%% file: main.tex
\newif\ifanonymousversion
\anonymousversionfalse 

\documentclass[11pt,conference]{IEEEtran}

\usepackage{graphicx}
\usepackage[bookmarks=false,hidelinks]{hyperref}
\usepackage[capitalise,noabbrev,nameinlink]{cleveref}
\usepackage{xurl}
\usepackage[keeplastbox]{flushend}

\newcommand{\nsf}[1]{\href{https://www.nsf.gov/awardsearch/showAward?AWD_ID=#1}{#1}}
\newcommand{\email}[1]{\href{mailto:#1}{#1}}

\date{}

\ifanonymousversion
\newcommand{\mythanks}[0]{}
\else
\IEEEoverridecommandlockouts
\newcommand{\mythanks}[0]{\thanks{
This work was supported in part by NSF grant \nsf{1901901}.
}}
\fi

\makeatletter 
\newcommand{\linebreakand}{%
 \end{@IEEEauthorhalign}
 \hfill\mbox{}\par
 \mbox{}\hfill\begin{@IEEEauthorhalign}
}
\makeatother 

\usepackage{xcolor}
\usepackage{soul}

\begin{document}

\title{Towards an Antivirus for Quantum Computers\mythanks{}}

\ifanonymousversion

 \author{Anonymous Author(s)\vspace{2.0cm}}

\else

\author{

\IEEEauthorblockN{Sanjay Deshpande}
\IEEEauthorblockA{\textit{Yale University} \\
New Haven, CT, USA \\
\email{sanjay.deshpande@yale.edu}}
\and
\IEEEauthorblockN{Chuanqi Xu}
\IEEEauthorblockA{\textit{Yale University} \\
New Haven, CT, USA \\
\email{chuanqi.xu@yale.edu}}
\and
\IEEEauthorblockN{Theodoros Trochatos}
\IEEEauthorblockA{\textit{Yale University} \\
New Haven, CT, USA \\
\email{theodoros.trochatos@yale.edu}}
\linebreakand
\IEEEauthorblockN{Yongshan Ding}
\IEEEauthorblockA{\textit{Yale University} \\
New Haven, CT, USA \\
\email{yongshan.ding@yale.edu}}
\and
\IEEEauthorblockN{Jakub Szefer}
\IEEEauthorblockA{\textit{Yale University} \\
New Haven, CT, USA \\
\email{jakub.szefer@yale.edu}}
}

\fi

\maketitle

\input{abstract}

\input{introduction}

\input{background}

\input{evaluation}

\input{antivirus}

\input{conclusion}

\bibliographystyle{IEEEtran}
\bibliography{bibliography}

\end{document}

%% file: abstract.tex
\begin{abstract}

Researchers are today exploring models for cloud-based usage of quantum computers where multi-tenancy can be used to share quantum computer hardware among multiple users. Multi-tenancy has a promise of allowing better utilization of the quantum computer hardware, but also opens up the quantum computer to new types of security attacks. As this and other recent research shows, it is possible to perform a fault injection attack using crosstalk on quantum computers when a victim and attacker circuits are instantiated as co-tenants on the same quantum computer. To ensure such attacks do not happen, this paper proposes that new techniques should be developed to help catch malicious circuits before they are loaded onto quantum computer hardware. Following ideas from classical computers, a compile-time technique can be designed to scan quantum computer programs for malicious or suspicious code patterns before they are compiled into quantum circuits that run on a quantum computer.  This paper presents ongoing work which demonstrates how crosstalk can affect Grover's algorithm, and then presents suggestions of how quantum programs could be analyzed to catch circuits that generate large amounts of crosstalk with malicious intent. 

\end{abstract}

%% file: introduction.tex
\section{Introduction}

Researchers across the world are today working towards building large-scale quantum computers, e.g., IBM most recently announced their $127$-Qubit Quantum processor called eagle in 2021~\cite{ibm_newsroom_2021}. In parallel research, several schemes are being proposed for sharing quantum computer hardware following ideas of multi-tenancy~\cite{10.1145/3410463.3414659, 10.1145/3352460.3358287, 9407180}. These schemes envision throughput optimization by allowing multiple users to run their circuits on the same quantum computer, but using different qubits. Although this idea has a good potential to improve utilization of quantum computer resources, it opens up quantum computers to security threats due to crosstalk errors~\cite{saki_abdullah}.

Crosstalk errors can occur when one or more qubits are activated, which adversely affects the state of other nearby qubits~\cite{osti_1513744}. If the former qubits are controlled by an attacker circuit, and the latter belong to a victim circuit, then the attacker can affect computational results of the victim~\cite{10.1145/3370748.3406570}.

As this work-in-progress proposes, it may be possible to achieve both multi-tenancy and security by introducing techniques to catch malicious quantum computer programs during compilation, before they are able run on the quantum computer.  In particular, quantum computer users today develop their designs in high-level programming languages or in quantum assembly language, which are then transpiled using tools such as Qiskit%
\footnote{Qiskit is an open-source software development kit for working with quantum computer programs, available at: \url{https://qiskit.org/}.}
into circuits that are scheduled and eventually loaded onto quantum computer hardware for execution. During this transpilation process, we propose that antivirus-like software can scan the user's programs and locate any malicious or suspicious code or circuit patterns. Based on existing work~\cite{10.1145/3370748.3406570} and our exploration, quantum {\tt CNOT} gates can be used to generate crosstalk errors. Following this, antivirus-like software can be used to scan if quantum programs contain suspicious patterns of {\tt CNOT} gates.

%% file: background.tex
\section{Background and Related Work}

Existing quantum computers are prone to different errors such as single-qubit and two-qubit gate errors, decoherence errors, crosstalk errors, and readout errors~\cite{Ding2020QuantumCS}. Out of these errors, crosstalk errors are of special interest as they lead to interesting security vulnerabilities~\cite{saki_abdullah}. The work from Ghosh et al.~\cite{10.1145/3370748.3406570} shows that the crosstalk errors could be used in a fault injection attack. It also showed how an adversary can launch a denial of service attack on a victim circuit using crosstalk errors.

To counter such threats threats, existing work proposes, for example, execution of the target circuit and performing measurements to determine if the target circuits was affected by crosstalk errors from a specific attacker circuit~\cite{Sarovar2020detectingcrosstalk}. In contrast, our work-in-progress proposes that malicious circuits could be caught at compile time using an antivirus-like approach, before any circuit is executed on a quantum~computer.

%% file: evaluation.tex
\section{Defining and Testing Potential Malicious Quantum Computer Circuits}

\begin{figure}[t]
 \centering
 \includegraphics[width=0.7\linewidth]{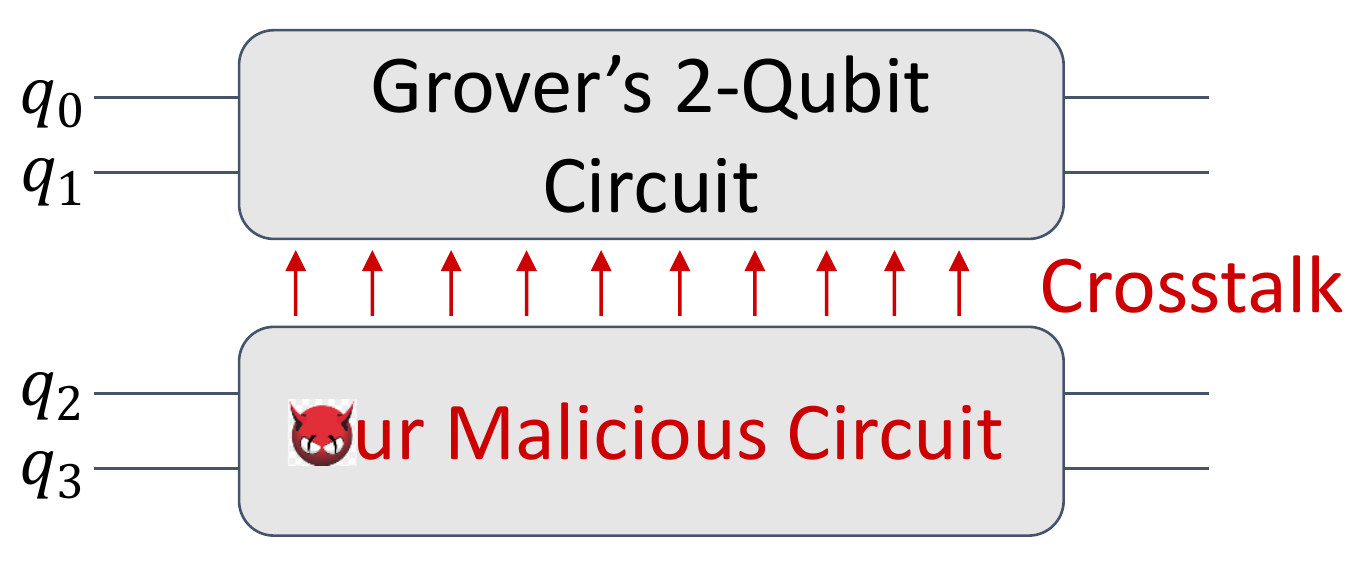}
 \caption{\small Block diagram of the experimental setup. A $5$-qubit quantum computer from IBM, namely the \texttt{ibmq\_lima}, is used to evaluate crosstalk effects of different attacker circuits on $2$-qubit Grover's algorithm. Note that the fifth qubit is unused in this example and only two qubits are needed for victim and two more for the attacker.}
 \label{fig:setup}
\end{figure}

To design an antivirus program, we first need to define what a virus is to quantum computers. We define a virus to be a malicious circuit that causes the crosstalk errors and affects the output fidelity of another target victim circuit. However, as with classical computer antivirus, the definitions of the virus can be updated and broadened over time as new understanding of potentially malicious circuits is~developed. As such, we propose to keep a dynamic library of virus quantum circuits. When a potential attacker circuit is evaluated, we label it as a virus circuit if it indeed causes crosstalk errors in an adjacent victim co-tenant. Patterns of such virus circuits can then be added to a database that the antivirus software uses when scanning for malicious attackers.

\subsection{Finding Malicious Circuit Types}
\label{sec:expsetup}

As a preliminary demonstration, we use the $5$-qubit IBM quantum computer \texttt{ibmq\_lima}~\cite{ibmquantum} for conducting our experiments. The attack example is shown in~\cref{fig:setup}. We use Grover’s $2$-qubit algorithm circuit~\cite{team_2021} as the target victim circuit which is fixed on qubits $q_0$ and $q_1$. For the malicious circuit we use qubits $q_2$ and $q_3$, and we define different series of {\tt CNOT} 
gates as potential malicious circuits, following preliminary work by Ghosh et al.'s work~\cite{10.1145/3370748.3406570} which used a series of {\tt CNOT} as an attack circuit to perform fault injection attack. 

We observe that without active attack, the output probability of the Grover's circuit is $P_{original} = 0.87$. 
Interestingly, when the attacker circuit is a pure sequence of {\tt CNOT} gates, we do not notice significant reduction in the $P_{original}$. The reason for that is, the transpile function when ran in optimization mode decomposes all {\tt CNOT} gates and converts them in to a delay value equivalent to all the {\tt CNOT} gates that are connected in series. If the number of {\tt CNOT} gates connected in series are even, the series of gates is converted into delay and scheduled at the beginning of the operation and if it is odd then it is converted in to delay and one {\tt CNOT} gate connected in series and scheduled at the beginning of the operation. Interestingly, this result is in contrast to~\cite{10.1145/3370748.3406570}, which did not seem to consider such existing transipler optimizations and only tested a pure sequence of {\tt CNOT}~gates.

\begin{figure*}[t]
 \centering
 \includegraphics[width=0.6\linewidth]{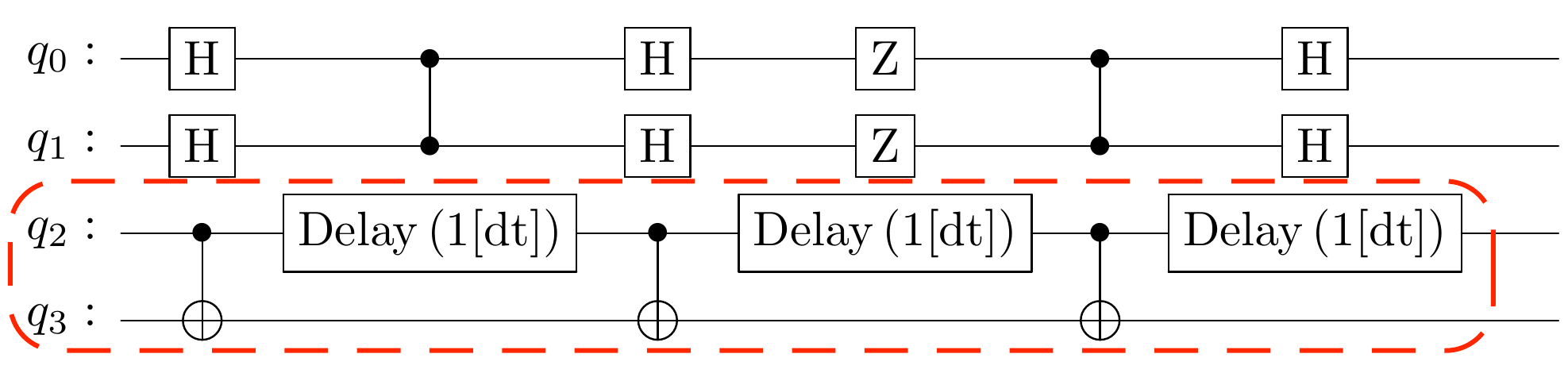}
 \caption{\small Circuit diagram for one possible attacker circuit with series of interleaved {\tt CNOT} and delay gates. The attacker circuit is outlined with a red dotted line, while the other circuit on qubits $q_0$ and $q_1$ is the victim Grover's circuit.}
 \vspace{1em}
 \label{fig:igcx}
\end{figure*}

In order to bypass the optimizations, we add delay gates~\cite{delay_team} with minimum delay in between each {\tt CNOT} gate in the malicious circuit as shown in the~\cref{fig:igcx}. And we noted that with this approach we were able to bypass the transpile optimizations and preserve the {\tt CNOT} gates in the malicious circuit. By running this circuit on the quantum device, the output probability value of target circuit is lower than $P_{original}$. 

We define two parameters to perform further analysis and find more malicious circuits. The first parameter is the delay value from the delay gate~\cite{delay_team}. The delay unit we use is $dt$ which translates to $2/5$ nanoseconds on the IBM quantum computers. The second parameter is $K$ that defines the number of gates in the malicious circuit. These malicious circuits are ran alongside Grover's 2-qubit algorithm in all cases. In the plots, the {\tt CNOT} gates are labeled as $CX$, while the other gates are the Pauli gates $X$, $Y$, and $Z$, and the $I$ gate.  The delay gates are labeled as $Delay(i)$, where $i$ is the delay of the delay gate given in units of $dt$.

Below we present the experiments performed for three cases of possible malicious circuits: series of {\tt CNOT} and delay gates in an interleaved fashion, as discussed before, series of only delay gates, and series of interleaved Pauli and delay gates. In all the cases we ran experiments by varying $K$ from $0$ to $5000$, but for brevity we only plot results for $K$ from $0$ to $300$.

\begin{figure}[t]
 \centering
 \includegraphics[width=1.0\linewidth]{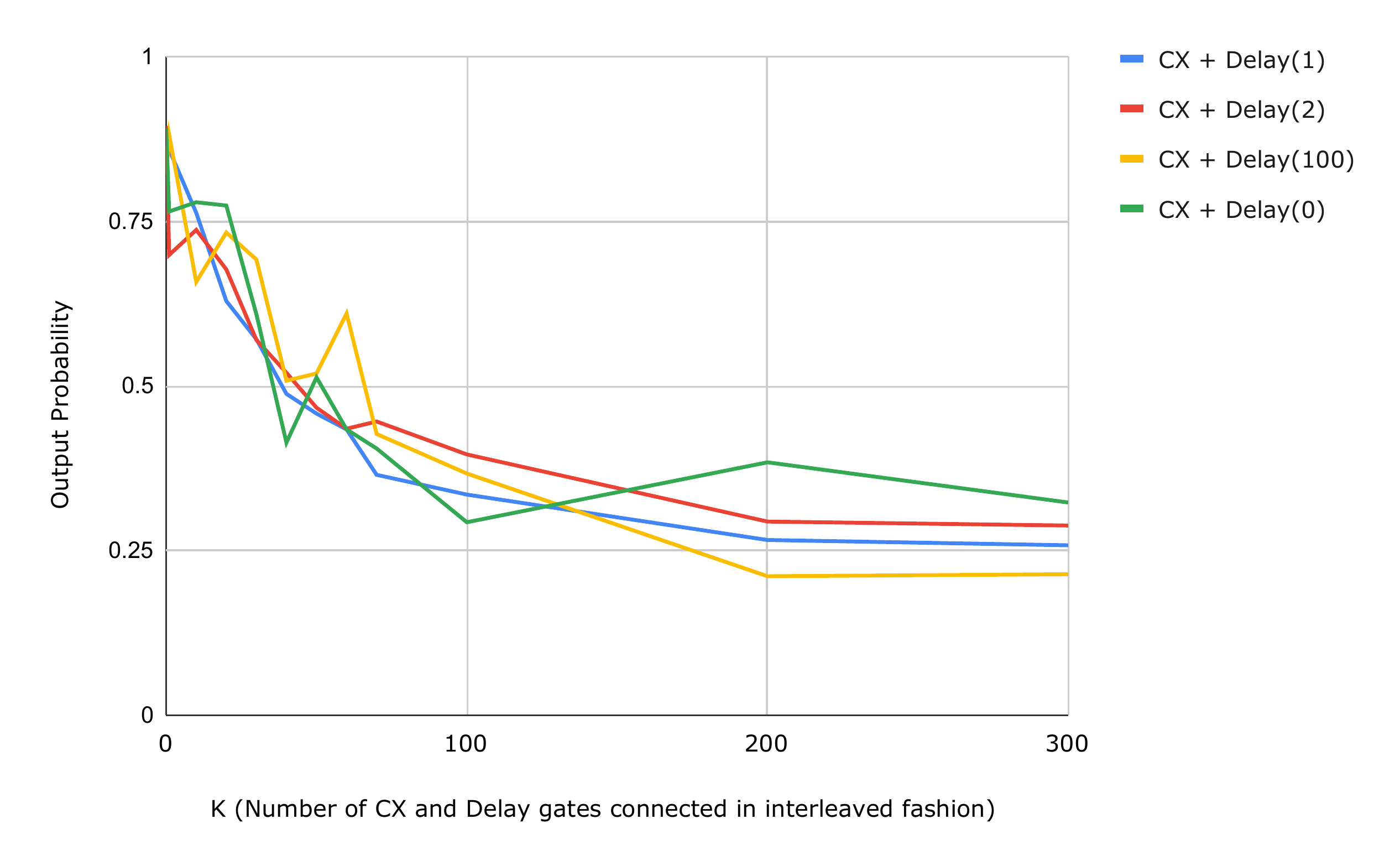}
 \caption{\small $K$ vs. Output Probability plot where malicious circuit is {\tt CNOT} (labled CX in the figure) and delay gates connected in an interleaved fashion.}
 \label{fig:k_vs_cxd}
\end{figure}

Considering {\tt CNOT} gates interleaved with delay gates, we ran experiments by fixing delay value to $1$ and varying $K$ and observe that as $K$ increase the output probability of target circuit decreases and it saturates around $0.2$ as shown in~\cref{fig:k_vs_cxd}. We ran similar experiments by fixing delay values to $2$ and $100$ and notice similar results. We also tried the delay value to be $0$ (i.e. no delay) and we notice that the transpiler optimization can be tricked even with the zero delay value and this has the same effect on the output probability as in cases of delay equal $1$, $2$, $100$, as shown in~\cref{fig:k_vs_cxd}.

\begin{figure}[t]
 \centering
 \includegraphics[width=1.0\linewidth]{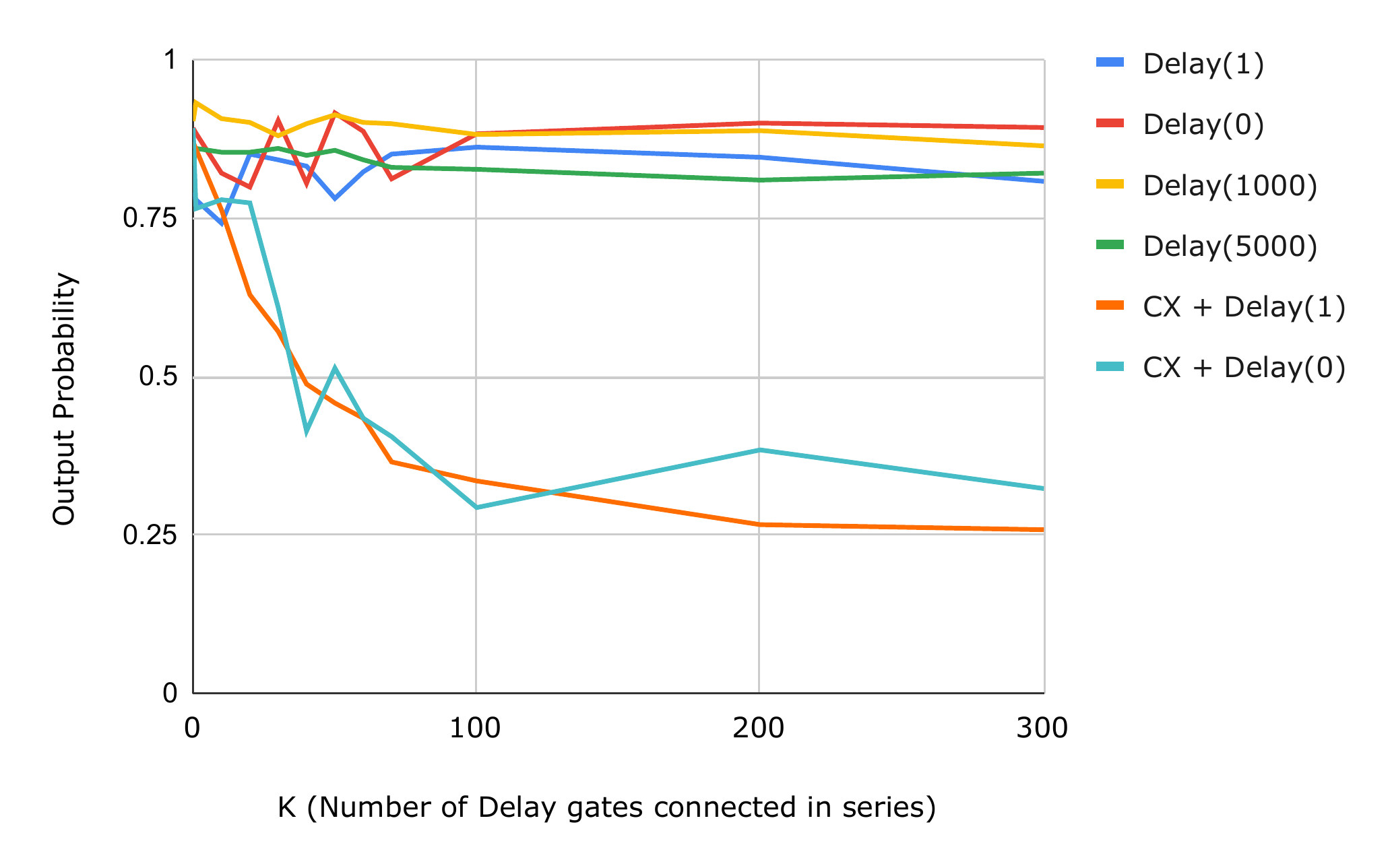}
 \caption{\small $K$ vs. Output Probability plot where malicious circuit is series of delay gates, delay length in units of $dt$ is given in the parenthesis in the legend; data for {\tt CNOT} gate experiments is shown for comparison.}
 \label{fig:k_vs_d}
\end{figure}

Considering only delay gates, we used the delay gates as malicious circuits to observe their effect on the target circuit. We ran experiments by varying $K$ and observed that there is not any significant change in the output probability of the target circuit as shown in~\cref{fig:k_vs_d}. The reason for this is that the transpile function identifies all the unwanted delays and schedules them at the beginning of the circuit and performs the operation of target circuit at the end. We also plot the results where the malicious circuit is series of {\tt CNOT} gates and delay gates in the same plot to provide a comparison. 

\begin{figure}[t]
 \centering
 \includegraphics[width=1.0\linewidth]{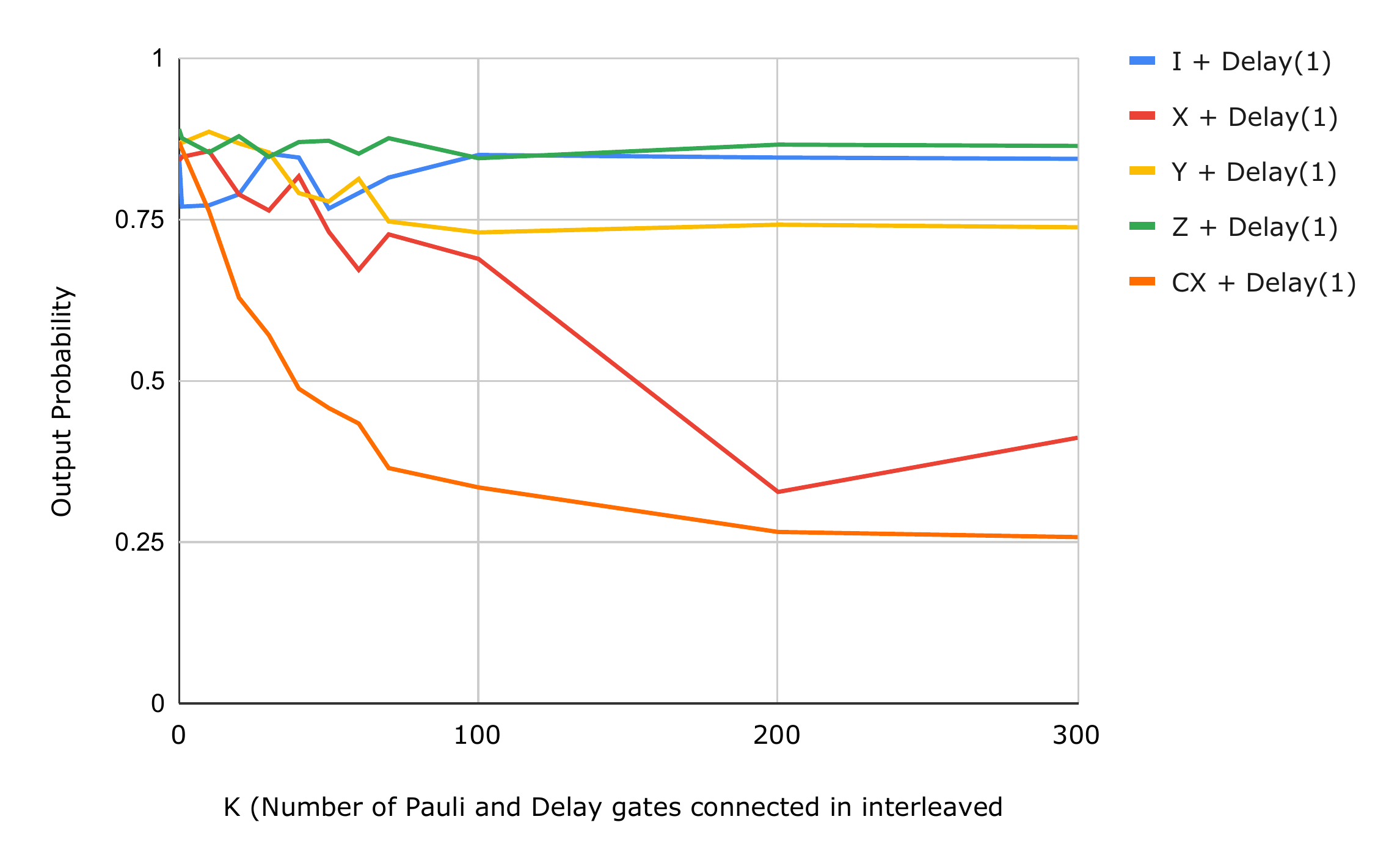}
 \caption{\small $K$ vs. Output Probability plot where malicious circuit is Pauli gates and delay gates connected in interleaved fashion; data for {\tt CNOT} gate experiments is shown for comparison.}
 \label{fig:k_vs_pd}
\end{figure}

Considering Pauli and delay gates, we used the series of Pauli and delay gates connected in interleaved fashion as malicious circuits. We ran experiments by varying $K$ and observed that there is some change in the output probability of the target circuit in case of $X$ and $Y$ gates (not as much as in case where malicious circuit is series {\tt CNOT} and delay gates) and we also observe there is not any significant change in output probability in case of $I$ and $Z$ gates as shown in~\cref{fig:k_vs_pd}. We plot the results where the malicious circuit is series of {\tt CNOT} gates and delay gates with delay value $1$ in the same plot to provide a comparison.

%% file: antivirus.tex
\section{Towards an Antivirus for Quantum Computers}
\label{sec:av}

The preliminary results show that the malicious circuits can be series of {\tt CNOT} gates interleaved with delay gates. They can also be circuits with interleaved gates such as Pauli $X$ and $Y$ gates, and thus can form a large or even an infinite malicious circuit set. On the other hand, pure delay gates, or $I$ and $Z$ gates do not produce noticeable crosstalk errors. Further, pure sequence of {\tt CNOT} gates is optimized away, so it does not induce crosstalk errors in the victim circuit. The former patterns that can be malicious could be used as the initial set of virus patterns that should be detected by the antivirus during the transpilation process.

\subsection{Quantum Computer Antivirus as a Qiskit Extension}

In order to detect and eliminate harmful circuits, the currently used Qiskit framework for developing quantum computer programs can be extended with the antivirus and pattern matching features.

As one possibility, an algorithm can be developed to count the number of appearances of each pattern in the quantum circuit. With a database of malicious patterns, the antivirus software can scan the user's code to find and count occurrences of the patterns. For example, as we have identified, when the value $K$ is low, there is low impact on the output probability of the victim Grover's circuit. As result if in a circuit being scanned there is only one {\tt CNOT} gate followed by a delay, i.e. $K=1$, then this is likely not a cirucit that could be malicious.  But if the count of {\tt CNOT} gate followed by a delay is greater than $5$ or $10$ this could be judged as possibly malicious circuit.

To realize the algorithm, both the quantum circuit and the patterns can be described by a set of Qasm instructions~\cite{qasm}. Therefore, the problem of quantum circuit matching can be reduced to find the instruction pattern in an instruction list. Compared with conventional string matching problem, the complexity of quantum circuit matching problem lies in the additional qubits (and possibly the classical bits and other parameters) for each instruction. Instructions that are not adjacent in the source code may still be gates in a series as well. There also may be multiple patterns dispersed among different qubits and in different order. Different patterns may cross with each other as well. All these challenges need to be addressed when creating a functional antivirus software for quantum computers.

%% file: conclusion.tex
\section{Conclusion}

This work proposes that multi-tenant quantum computers can be protected by developing mechanisms for scanning user's circuits for malicious patterns. By using an antivirus for preventing attackers from instantiating circuits which could be causing harmful crosstalk errors, secure multi-tenant cloud-based quantum computers can be realized.